\documentclass[doublecol,10pt]{epl2}
\usepackage{amsmath}
\usepackage{makeidx}
\usepackage{amsfonts}
\usepackage{graphicx}
\usepackage[ansinew]{inputenc}
\usepackage[usenames,dvipsnames]{pstricks}
\usepackage{subfigure}
\usepackage{epsfig}
\usepackage{pst-grad}
\usepackage{pst-plot}
\usepackage[brazilian,english]{babel}
\usepackage[colorlinks,hyperindex]{hyperref}

\hypersetup{
colorlinks,
citecolor=blue,
linkcolor=blue,
urlcolor=black,
}

\makeindex
\shorttitle{False Vacuum Transitions - Analytical Solutions and Decay Rate Values} 
\shortauthor{R. A. C. Correa \etal}
\institute{                    
  \inst{1} CCNH, Universidade Federal do ABC, 09210-580, Santo Andr\'e, SP,
Brazil\\
  \inst{2} DF, Instituto Tecnol\'ogico de Aeron\'autica, 12228-900, S\~ao Jos\'e dos
Campos, SP, Brazil\\
\inst{3} CMCC, Universidade Federal do ABC, 09210-580, Santo Andr\'e, SP,
Brazil
}
\abstract{
In this work we show a class of oscillating configurations for the evolution
of the domain walls in Euclidean space. The solutions are obtained
analytically. Phase transitions are achieved from the associated fluctuation determinant, by the decay rates of the false vacuum.}

\begin{document}

\title{False Vacuum Transitions - Analytical Solutions and Decay Rate Values}
\author{R. A. C. Correa\inst{1} \and P. H. R. S. Moraes\inst{2} \and %
Rold\~ao da Rocha\inst{3}}
\maketitle

\section{Introduction}

The interest in studying nonlinear systems has gradually grown since the
1950s, due to the fact that nonlinearity is present in many different areas
of science, including particle physics, plasma physics, cosmology and field
theory, among others. Nonlinear systems, particularly those having solitonic
excitations, play a prominent role in the modelling of several physical,
chemical and biological systems as well. For instance, in the electric
conductivity of organic materials, for which polarons and other polymer
chain solitons provide conducting polymers \cite{1}. Moreover, solitons are
also related to electric conduction through DNA molecules \cite{2}. In
particular, solitons play a fundamental role in field theory, encompassing
applications from particle physics to condensed matter. Non-abelian solitons
are specially important in gauge theories of elementary particle physics 
\cite{3}. Gravitating non-Abelian solitons were firstly discussed in the
context of four-dimensional Einstein-Yang-Mills theory \cite{6}. In the
context of string theory, supersymmetric solitons further play an important
role in the study of the non-perturbative sector and in understanding string
dualities. In condensed matter, the homotopy classification of finite energy
configurations can be realized by similar forms as can be accomplished in
field theory, when the solitary-wave solutions of SU(2) gauge theory are
classified likewise.

When a model with a field dependent potential, which has two or more
degenerate minima, is taken into account, different ground states arise at
different portions of the space. For instance, the so called domain walls
can be described \cite{gremm,dwolfe}, connecting different portions of the
space where the field has different values for the potential degenerate
minima, even further in asymmetric scenarios \cite{Bazeia:2013usa}. In other
words, the field configuration interpolates between the potential minima.
Solitons can describe such field configurations presenting a localized and
shape-invariant aspect, and have a finite energy density \cite{Rajaraman}.
The presence of those configurations is well understood in a wide class of
models, that encompass monopoles, textures, strings and kinks as well \cite%
{Vilenkin}.

In a seminal work by Coleman, a classical field theory was presented for the
study of the false vacuum decay in theories involving asymmetrical
potentials by analysing a kind of $\lambda \phi ^{4}$ asymmetric model. In
this theory, the relative minimum corresponds to the false vacuum, whereas
the absolute minimum corresponds to the true one. After that, Callan and
Coleman presented the associated quantum corrections for the theory \cite%
{Callan:1977pt}. These preliminary results paved an increasing interest in
cosmology, providing models where the scalar field potential driving
inflation has low energy minima, also known as false vacua. The absolute
minimum of the energy density corresponds to the so called true vacuum state
of the universe.

An important application of this theory is the process that involves phase
transitions in statistical mechanics. In this case, inside the false vacuum,
it occurs the formation of a bubble of true vacuum which initiates the
decay. From a quantum mechanical point of view, it corresponds to a
tunnelling probability from a false vacuum to the true one. Moreover, the
theory of vacuum decay has a considerable physical importance and can
include the effects of gravity \cite{coleman/1980}. In fact, in a
cosmological perspective, the early universe had an extremely high energy
density in a region of false vacuum. As it expands and cools down, it passes
through a phase transition towards the true vacuum \cite{coleman/1977}.

A wide class of problems involving phase transitions are supported by the $%
\phi ^{4}$ theory. However, due to the nonlinearity of this theory it is
very rare to analytically approach the problem. Recently an interesting
model \cite{stavros} was proposed, named asymmetrical double-quadratic
model. This model is similar to the asymmetric $\lambda \phi ^{4}$
characterized by the following potential 
\begin{equation}
V(\phi )=\frac{1}{2}\phi ^{2}-\left\vert \phi \right\vert -\epsilon \phi +%
\frac{1}{2}(\epsilon -1)^{2}.\   \label{pot}
\end{equation}

A similar model has been extensively studied in the literature in several
contexts. As an example, oscillons configurations \cite{gleiser/1994}, which
are time-dependent and long living solutions, were obtained in the presence
of the so-called signum-Gordon model \cite{arodz/2008}.

On the other hand, some years ago the Asymmetrical Double-Quadratic (ADQ)
model allowed the attainment of oscillating configurations in Minkowski
space-time \cite{dutaraccorrea1}. In fact, oscillating configurations were
shown to be responsible for a delay in the transition from a false to a true
vacuum. Besides, a connection with phase transitions, which occur in
ferromagnetic materials, was presented. In addition, fermions bound states
were proposed in a background of static solutions provided from ADQ model 
\cite{{dutraraccorrea2}}.

This paper is organized as follows: in Section 2 we present the asymmetrical
model to be analysed and the classical field configurations. In Section 3 we
obtain the analytical solutions for the scalar field. In Section 4 we
calculate the respective decay rates. 
In Section 5 we present the conclusions.

\section{Classical field configurations}

\label{cfc}

\textcolor{black}{The so called
Generalized Asymmetrical Doubly Quadratic model (GADQM) was studied in  \cite{dutraraccorrea2}, corresponding 
to a generalized version of the ADQ model. The advantage of the GADQM is
that the vacua can be chosen to represent a slow-roll potential, yielding inflaton fields which are important in cosmological
inflationary scenarios. The authors in \cite{dutraraccorrea2} considered the
problem of fermion bound-states and zero modes in the background of kinks of
the GADQM. Motivated by that work, Brito, Correa, and Dutra \cite{dutracorreagustavo}  showed an approach to construct nonlinear systems with analytical multikink profile configurations. A prominent consequence of this work is the possibility of the resulting
field configurations can be applied to the study of problems of condensed
matter, cosmology, and braneworld scenarios. Thus, in order to employ this model, that moreover is exactly
solvable, we will} study the model whose potential is given by \cite%
{dutraraccorrea2, dutracorreagustavo} 
\begin{subequations}
\begin{eqnarray}
&&\!\!\!\!\!\!\!\!\!\!\!\!\!\!\!\!\!\!V_{1}(\phi )=\frac{\lambda _{1}}{2}%
\left[ \left( \phi +\frac{3a}{2}\right) ^{2}+b_{1}\right] ,\;\text{ }-\infty
<\phi \leq -a\,,  \label{1} \\
&&  \notag \\
&&\!\!\!\!\!\!\!\!\!\!\!\!\!\!\!\!\!\!V_{2}(\phi )=\frac{\lambda _{2}}{2}%
\left[ \left( \phi +\frac{a}{2}\right) ^{2}+b_{2}\right] ,\text{ }-a\leq
\phi \leq 0\,, \\
&&  \notag \\
&&\!\!\!\!\!\!\!\!\!\!\!\!\!\!\!\!\!\!V_{3}(\phi )=\frac{\lambda _{3}}{2}%
\left[ \left( \phi -\frac{a}{2}\right) ^{2}+b_{3}\right] ,\text{ }0\leq \phi
\leq a\,,  \label{1.1.1} \\
&&  \notag \\
&&\!\!\!\!\!\!\!\!\!\!\!\!\!\!\!\!\!\!V_{4}(\phi )=\frac{\lambda _{4}}{2}%
\left[ \left( \phi -\frac{3a}{2}\right) ^{2}+b_{4}\right] \,,\text{ }a\leq
\phi <\infty \,,  \label{1.123}
\end{eqnarray}%
\noindent where $\lambda _{i+1}=\lambda _{1}\left( \frac{a^{2}+4b_{1}}{%
a^{2}+4b_{2}}\right) \,$ and $i=1,2,3\,.$ Fig. 1 depicts this potential,
presenting four asymmetric vacua located at $\pm 3a/2$, $\pm a/2$. 
\textcolor{black}{Moreover, we can see in Fig. 1  that the potential has three false vacua and a
true vacuum. Thus, according to the pioneering work of Coleman \cite{coleman/1977}, the false vacua can decay, where phase
transitions play an important role in various phenomena. For
example, it occurs in the nucleation process of statistical physics  \cite{vac}, in  cosmological
contexts \cite{andrei}, and in the Weinberg-Salam model  \cite{ws} as well.}

\begin{figure}[h]
\begin{center}
\includegraphics[width=7cm]{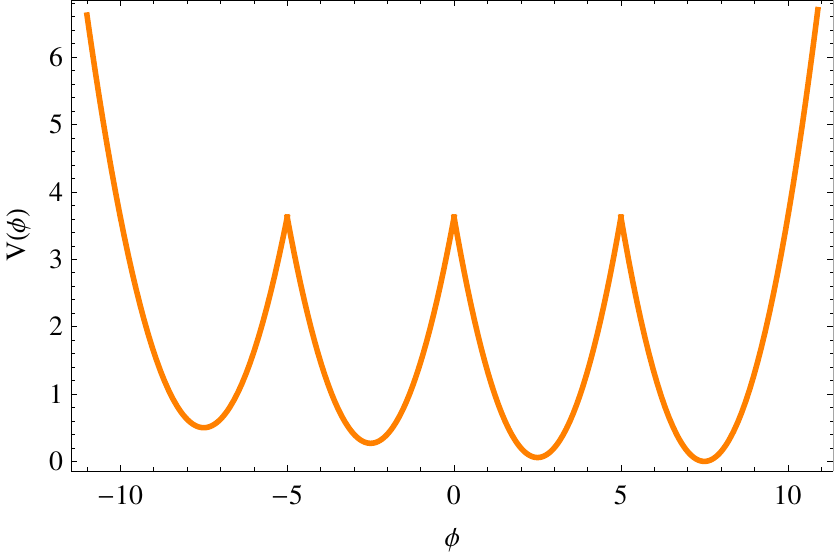} \includegraphics[width=7.2cm]{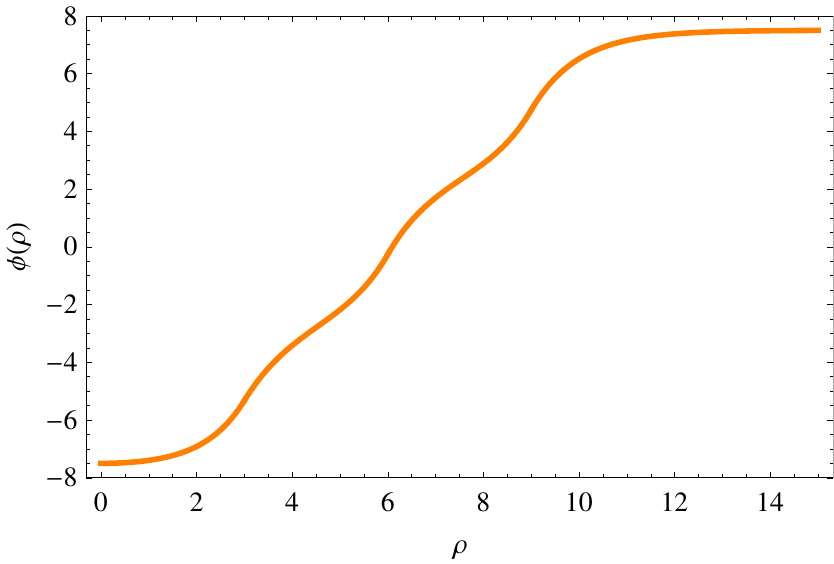}
\end{center}
\caption{Potential of the model (top)\ and corresponding field configuration
(bottom) for $\protect\nu =1$, $\protect\lambda _{1}=1$, $a=5$, $b_{1}=1$, $%
b_{2}=0.5$, $b_{3}=0.1$ and $b_{4}=0$.}
\end{figure}

In this case it is possible to have a phase transition between vacua. It is
important to remark that a class of configurations in 1, 2 and 3 dimensions,
in a similar potential, was presented in refs. \cite%
{dutaraccorrea1,dutracorreagustavo}, proposing the existence of kink
configurations in the presence of the ADQ model. 
\textcolor{black}{At this point, it is
important to point out that in refs. \cite{dutaraccorrea1,dutracorreagustavo}
the spacetime used by the authors is Minkowskian, and the field
configurations correspond to the static case.} Here we will work in a single
scalar field theory in four-dimensional spacetime, which has the following
Euclidean action

\end{subequations}
\begin{equation}
S_{E}[\phi ]=\int d^{4}x\left[ \frac{1}{2}\nabla _{\mu }\phi \nabla ^{\mu
}\phi +V(\phi )\right] ,  \label{2}
\end{equation}%
\noindent where $V(\phi )$ is given by the potential shown in Eqs.(\ref{1})-(%
\ref{1.123}). 
\textcolor{black}{Note that the Euclidean theory is left invariant by $O(4)$
transformations. Moreover, the Euclidean system is not similar as the
Minkowskian system, since $(x_{0}^{2}-\vec{x}^{2})$ for Minkowski coordinates is invariant, whereas the Euclidean version is provided by $\sum\limits_{\mu
=1}^{4}(x^{\mu })^{2}$ .} 
\textcolor{black}{Hence, the two systems are
different, although the tunnelling processes in field theory are described
by the field equations in Euclidean spacetime}. Now, from the variation of
the above action, the corresponding Euclidean equation of motion is provided
by 
\begin{equation}
\left( \frac{\partial ^{2}}{\partial \tau ^{2}}+\sum_{i=1}^{N}\partial
_{i}\partial _{i}\right) \phi =\frac{\partial V}{\partial \phi },  \label{3}
\end{equation}

\noindent where $N$ represents the number of dimensions. 
\textcolor{black}{Moreover, as a
consequence of the Euclidean theory, we can observe in the above equation an
elliptic differential operator, which opens new  possibilities to find a different
set of real non-singular solutions.} For the model to be analyzed, Eq.(\ref%
{3}) reads 
\begin{equation}
\left. \frac{d^{2}\phi _{j}(\rho )}{d\rho ^{2}}\!+\!\frac{N}{\rho }\frac{%
d\phi _{j}(\rho )}{d\rho }\!-\!\lambda _{j}\phi _{j}(\rho )=\lambda
_{j}A_{j},\right. j=1,\ldots ,4\,,  \label{eq.22}
\end{equation}%
\noindent with $\rho =\sqrt{|\vec{x}|^{2}+\tau ^{2}}$, wherein $\vec{x}$
represents the three spatial coordinates and $\tau $ the Euclidean time, $%
A_{1}=3a/2$, $A_{2}=a/2$, $A_{3}=-a/2$ and $A_{4}=-3a/2$.

Now, in order to solve Eq.(\ref{eq.22}), the transformations $\phi _{j}(\rho )=-A_{j}+\varphi _{j}(\rho )$ in
the fields $\phi _{j}$ are performed. Thus, it is not difficult to conclude that 
\begin{equation}
\frac{d^{2}\varphi _{j}(\rho )}{d\rho ^{2}}+\frac{N}{\rho }\frac{d\varphi
_{j}(\rho )}{d\rho }-\lambda _{j}\varphi _{j}(\rho )=0.  \label{5.11}
\end{equation}

\section{Analytical solutions}

\label{as}

There are many possible ways to solve the above equation, each of them with
advantages and disadvantages. We will work with the one which makes possible
to find real solutions. Thus, in order to solve the above equation, we
choose 
\begin{equation}
\varphi _{j}(\rho )=\rho ^{\nu }\Phi _{j}(\rho ).
\end{equation}%
Hence, the remaining function $\Phi _{j}(\rho )$ satisfies the Bessel
equation 
\begin{equation}
\rho ^{2}\frac{d^{2}\Phi _{j}(\rho )}{d\rho ^{2}}+\rho \frac{d\Phi _{j}(\rho
)}{d\rho }-(\nu ^{2}+\lambda _{j}\rho ^{2})\Phi _{j}(\rho )=0,
\end{equation}%
\noindent where $\nu \equiv (N-1)/2$. If the variable $Z=\rho \sqrt{\lambda
_{j}}$ is taken into account, it yields the well known solution 
\begin{equation}
\Phi _{j}(Z)=a_{j}K_{\nu }(Z)+b_{j}I_{\nu }(Z),  \label{7}
\end{equation}
where $K_{\nu }(Z)$ and $I_{\nu }(Z)$ are modified Bessel functions of order 
$\nu $. Therefore, the complete solution has the form 
\begin{equation}
\phi _{j}(\rho )=-A_{j}+\rho ^{\nu }[a_{j}K_{\nu }(\rho \sqrt{\lambda _{j}}%
)+b_{j}I_{\nu }(\rho \sqrt{\lambda _{j}})].  \label{9.11}
\end{equation}%
Now we search for configurations where four different regions do exist,
wherein the field is connected. At this point it is important to remark that
the solutions in each region must be continuous through the $\rho $ axis,
what leads to $\phi _{1}(\rho _{1})=\phi _{2}(\rho _{1})$, $\phi _{2}(\rho
_{2})=\phi _{3}(\rho _{2})$ and $\phi _{3}(\rho _{3})=\phi _{4}(\rho _{3})$.
On the other hand, we are looking for the case where $\phi _{1}(\rho
=0)=-3a/2$ and $\phi _{4}(\rho \rightarrow \infty )=3a/2$, which results in 
\begin{eqnarray}
&&\!\!\!\!\!\!\!\!\!\!\!\!\!\phi _{1}(\rho)=-\frac{3a}{2}+\rho ^{\nu
}b_{1}I_{\nu }(\rho \sqrt{\lambda _{j}}),\text{ \ }0\leq \rho \leq \rho _{1}
\label{9} \\
&&\!\!\!\!\!\!\!\!\!\!\!\!\!\phi _{2}(\rho )=-\frac{a}{2}\!+\!\rho ^{\nu
}[a_{2}K_{\nu }(\rho \sqrt{\lambda _{j}})\!+\!b_{2}I_{\nu }(\rho \sqrt{%
\lambda _{j}})],\rho _{1}\leq \rho \leq \rho _{2}  \notag \\
&&\!\!\!\!\!\!\!\!\!\!\!\!\!\phi _{3}(\rho ) =\frac{a}{2}\!+\!\rho ^{\nu
}[a_{3}K_{\nu }(\rho \sqrt{\lambda _{j}})\!+\!b_{3}I_{\nu }(\rho \sqrt{%
\lambda _{j}})],\text{ }\rho _{2}\leq \rho \leq \rho _{3}  \notag \\
&&\!\!\!\!\!\!\!\!\!\!\!\!\!\phi _{4}(\rho ) =\frac{3a}{2}+\rho ^{\nu
}a_{4}K_{\nu }(\rho \sqrt{\lambda _{j}}),\text{ \ }\rho _{3}\leq \rho
<\infty .  \label{12}
\end{eqnarray}

To determine the constants $a_{j}$ and $b_{j}$ we use the condition of
continuity of the function $\phi _{j}$ and of its derivative, in each point
of the region $\rho $. Here, it is worth to remark that these conditions
generate a set of six equations which allows the determination of the values
of these constants. {For instance, we show the profile of the field
configuration in Fig. 2 for a given set of values of the parameters of the
model.} 
\begin{figure}[h]
\begin{center}
\includegraphics[width=8cm]{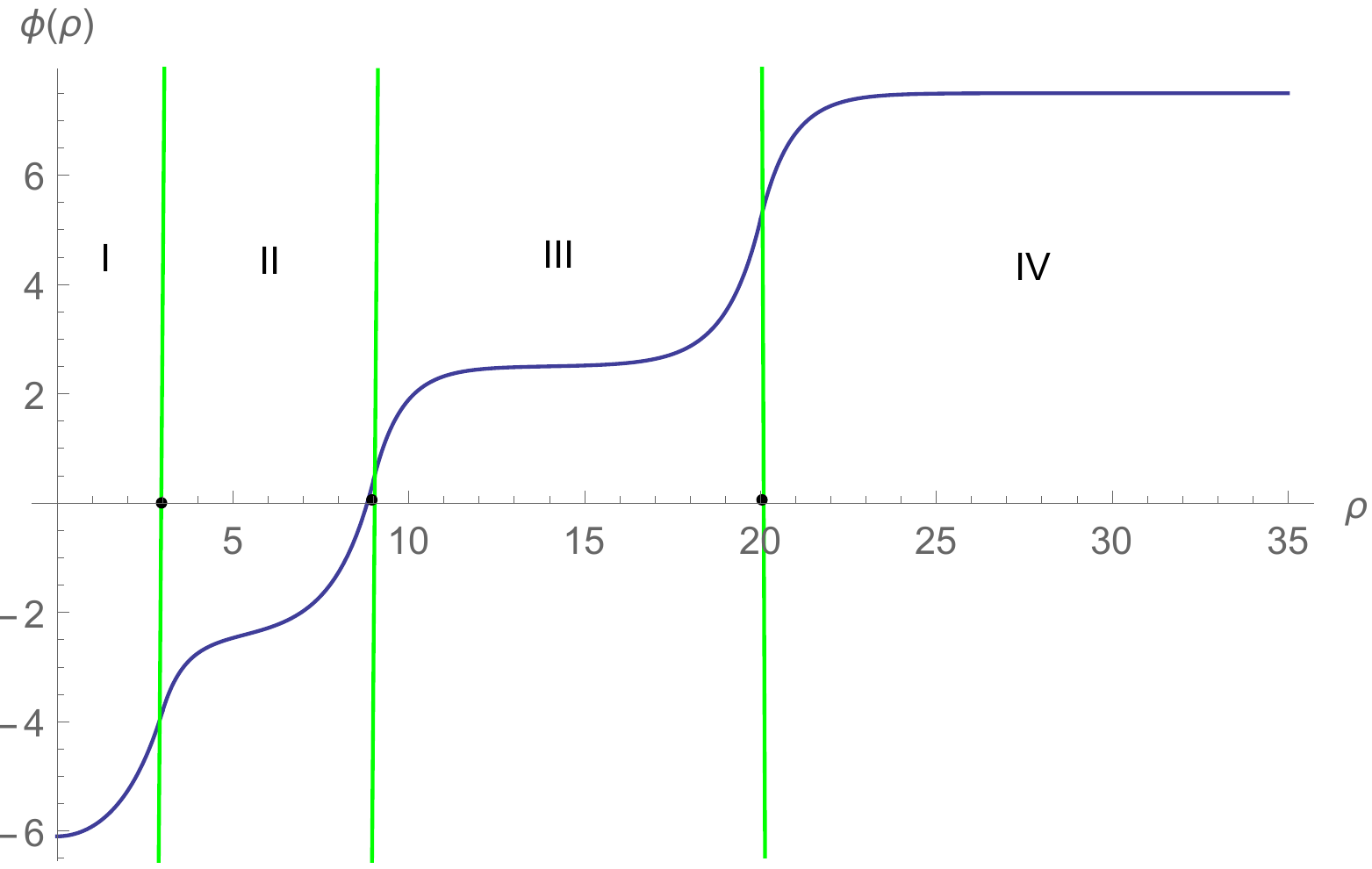}
\end{center}
\caption{Field configuration for $\protect\nu =1$, $\protect\lambda _{1}=1$, 
$a=5$, $b_{1}=1$, $b_{2}=0.5$, $b_{3}=0.1$, $b_{4}=0$, $\protect\rho _{1}=3$%
, $\protect\rho _{2}=9$ and $\protect\rho _{3}=20$.}
\label{Figure2}
\end{figure}
Moreover, it is necessary to impose that the first derivative of the field
configuration is continuous at the intermediate points. Such a constraint
produces transcendental equations which restrict the distances among the
roots of the scalar field $\phi$.

\section{First-order phase transitions}

\label{fopt}

{As argued in the seminal work by Coleman \cite{coleman/1977}, a field
theory with false vacuum allows to understand a large number of phenomena
from condensate matter physics to cosmology. The key for this understanding
comes from the decay probability per unit volume, which is} %
\textcolor{black}{named} 
\textcolor{black}{ decay rate. 
In particular, in a cosmological context, we can suppose that when the
universe had been created, it was far from any vacuum state. As it has
expanded and cooled down, it first evolved to a false vacuum instead of the
true one. Thus, in such a scenario}, 
\textcolor{black}{when the cosmological
time elapses,} 
\textcolor{black}{the universe
should finally be settled in the true vacuum state. Therefore,} in this
section we calculate the decay rate of each phase transition that occurs
from the false vacuum to the true vacuum.

\textcolor{black}{In fact,  let us denote as usual the two stable static
states by $\phi=\phi_{fv}$ (the false vacuum). Since $\phi_{fv}$ is unstable, its
energy $E(\phi_{fv})$ must present an imaginary part responsible for its decay
width, as Im $E(\phi_{fv}) = -\Gamma(\phi_{fv})/2$. In a weakly coupling
framework, the width $\Gamma(\phi_{fv})$ is exponentially small. The energy of
the false vacuum reads a functional integral of type $\int \mathcal{D}\phi
e^{-S[\phi]}$, where $S_E[\phi]$ regards Eq.(\ref{2}). Hence the width
$\Gamma$, which corresponds to the decay rate, is determined by the
contribution of a non-trivial saddle point, named the bounce $\phi_B$, of
the Euclidean action $S_E[\phi]$} \cite%
{Callan:1977pt,coleman/1980,coleman/1977}.

It is important to remark that the decay rate represents the tunnelling
probability from the false vacuum to the true one, per unit time per unit
volume $U$, in such a way that in the semiclassical limit it assumes the
form 
\begin{equation}
\Gamma /U\sim K\exp (-S_{E}[\phi _{cl}]),\ 
\end{equation}%
\textcolor{black}{where the prefactor term $K$, which corresponds to quantum fluctuations
with respect to the classical solution at the bounce field, is a coefficient that depends on the detailed
form of $V(\phi)$.} {Indeed, Coleman and Callan proved that the contribution
of the bounce to the energy of the false vacuum reads \cite{Callan:1977pt} 
\begin{equation}
E_{cl} \!=\! -i\frac{U}{2}\left(\!\frac{S_{E}[\phi _{cl}]^{2}}{4\pi ^{2}}\!\right)e^{{S_{E}}[\phi _{cl}]}\sqrt{\left\vert \frac{\det [-\partial ^{\mu
}\partial _{\mu }+V^{{\prime \prime }}(\phi _{fv})]}{{\mathring{\det }}[-\partial ^{\mu }\partial _{\mu }+V^{{\prime \prime }}(\phi _{cl})]}\right\vert }.  \notag  \label{15aa}
\end{equation}}{In the above equation, the prime denotes the derivative with respect to
the argument, $\phi _{fv}$ is the false vacuum, and $\phi _{cl}$ is a
classical solution {associated to the bounce $\phi_B$.} Furthermore,} the
ringed determinant in Eq.(\ref{15}) denotes that the zero modes of the
operator $-\partial _{\mu }\partial ^{\mu }+V^{{\prime \prime }} $,
corresponding to the translational invariance about the bounce location, are taken out.

\textcolor{black}{More precisely, the contribution of $\phi_{fv}$  is determined by a Gaussian integral over the perturbations about the bounce, namely  $\phi(x) = \phi_{fv} + \eta(x)$. Hence (\ref{2}) yields the fluctuations $S(\eta) = \int d^4x\;\frac12\eta\left[-\partial^2_\mu+V^{\prime\prime}(\phi_{fv})\right]\eta$. Ref. \cite{rubakov} introduces the eigenfunctions $\eta_\lambda$ as 
 \begin{equation}\label{bounce1}[-\partial^2_\mu+V^{\prime\prime}(\phi_{fv})]\eta_\lambda=\lambda\eta_\lambda\,,\end{equation}\noindent 
by the splitting $\eta(x) = \sum_\lambda\;a_\lambda\eta_\lambda(x)$. Hence the contribution of the false vacuum to the integral $\int \mathcal{D}\phi
e^{-S[\phi]}$ reads $I_0 = \int\prod_\lambda\left[{da_\lambda}\right]\exp\left[-\frac12\sum_\lambda \lambda a_\lambda^2\right]$, or equivalently, to $I_0=\prod_\lambda \lambda^{-1/2}$. The quantity $\prod_\lambda \lambda$ can be interpreted as the determinant of the operator $-\partial_\mu\partial^\mu+V^{\prime\prime}(\phi_{fv})$, since it corresponds to the product of its eigenvalues. Thus $I_0 = \det [-\partial ^{\mu
}\partial _{\mu }+V^{{\prime \prime }}(\phi _{fv})]$.
Moreover, Eq. (\ref{bounce1}) can also present zero eigenvalues (modes). In fact, the center of the bounce may be located at any point in Euclidean space-time, leading to the existence of four zero modes around the bounce, determined by $\eta_\mu = S^{-1/2}\partial_\mu\phi_B(x), \mu = 0,1,2,3$ \cite{rubakov}. By treating the zero modes in the same way as non-zero modes, then the corresponding integral would diverge \cite{rubakov}, what imposes that the zero modes are dismissed from the calculations in the previous formula for $E_{cl}$, accordingly.}

Alternatively, the bounce action $S_{E}[\phi _{cl}]$ is evinced
in the right-hand side of Eq.(\ref{15}). {Hence, in general, the above
mentioned pre-factor reads} \cite{dunne/2006} 
\begin{equation}
K=\frac{S_{E}[\phi _{cl}]^{2}}{4\pi ^{2}}\sqrt{\left\vert \frac{\det
[-\partial ^{\mu }\partial _{\mu }+V^{{\prime \prime }}(\phi _{fv})]}{{%
\mathring{\det }}[-\partial ^{\mu }\partial _{\mu }+V^{{\prime \prime }%
}(\phi _{cl})]}\right\vert }.  \label{15}
\end{equation}
Now, in order to evaluate {the decay rate per unit volume and unit time} $%
\Gamma /U$, firstly we need to find the classical solution $\phi _{cl}$.
Thus, from the Euclidean action (\ref{2}), we have the following equation of
motion 
\begin{equation}
\left( \frac{d^{2}}{d\rho ^{2}}+\frac{N}{4}\frac{d}{d\rho }\right) \phi
_{cl}-V^{\prime }(\phi _{cl})=0.
\end{equation}

{Here, for the model here presented by Eqs. (\ref{1}-\ref{1.123}),} $\phi
_{cl}$ represents the classical solution for each region. Thus, using Eqs. (%
\ref{9}-\ref{12}) we can obtain the correct form of the Euclidean action,
which is fundamental for the decay rate calculation. 
\textcolor{black}{For simplicity, we can rewrite the Euclidean
action in the following compact form\begin{equation}
S_{E}[\phi _{cl}]=\sum_{j=1}^{4}S_{E}^{(j)},  \label{ecl1}
\end{equation}
} \noindent where 
\begin{equation}
\!\!\!\!\!\!S_{E}^{(j)}\!=\!\frac{2\pi ^{\nu\! +\!1}}{\Gamma (\nu\! +\!1)}%
\int_{\rho _{j-1}}^{\rho _{j}}\!\!\rho ^{2\nu +1}\left[ \frac{1}{2}\left( 
\frac{d\phi _{j}}{d\rho }\right) ^{2}+V_{j}(\phi _{j})\right] d\rho ,
\label{rel}
\end{equation}

\noindent with $\rho _{0}=0$ and $\rho _{4}=\infty $. In fact, both the
coefficients $K$ and the exponential factor explicitly depend upon the
action. Hence, for the model here analysed, the decay rate is given by the
transition from the false vacuum, located at $\phi _{fv}^{(1)}=-3a/2$, to
the true vacuum, at $\phi _{tv}^{(4)}=3a/2$. Note that before reaching the
point $\phi _{tv}^{(4)}$, the decay process rolls through the other two
metastable potentials, located at $\phi _{fv}^{(2)}=-a/2$ and $\phi
_{fv}^{(3)}=a/2$. Hence, the decay rate from the false vacuum to the true
vacuum becomes 
\begin{equation}
\Gamma /U=\sum_{j=1}^{3}(\Gamma _{j}/U_{j}),
\end{equation}%
with the following definition%
\begin{eqnarray}
\frac{\Gamma _{j}}{U_{j}} &:&=\frac{1}{(2\pi )^{2}}\left(
\sum_{n,m=1}^{4}S_{E}^{(n)}S_{E}^{(m)}\right) \left[ \left\vert \frac{\det (%
\mathcal{M}_{(j)})}{\det (\mathcal{M}_{(j)}^{0})}\right\vert \right] ^{-1/2}
\notag \\
&&\times \prod\limits_{q=1}^{4}\exp \left( -S_{E}^{(q)}\right) ,  \label{dqg}
\end{eqnarray}%
\noindent where we used the relation in Eq. (\ref{rel}). Moreover, we employ
the fluctuation operator $\mathcal{M}_{(j)}$ in the background of the
classical solution and its counterpart $\mathcal{M}_{(j)}^{0}$ in the false
vacuum, \textcolor{black}{which are given by%
\begin{eqnarray}
\mathcal{M}_{(j)} &=&-\partial ^{\mu }\partial _{\mu }+V_{j+1}^{{\prime
\prime }}(\phi ^{(j+1)}), \\
\mathcal{M}_{(j)}^{0} &=&-\partial ^{\mu }\partial _{\mu }+V_{j}^{{\prime
\prime }}(\phi ^{(j)}),
\end{eqnarray}}
 \noindent \textcolor{black}{where 
$
\phi ^{(a)} \rightarrow \phi _{fv}^{(a)}$, for $a=1,2,3,4$.} \textcolor{black}{Now, it is convenient to use the fact that we have the $O(4)$ spherical
symmetry, in such a way that the operators $\mathcal{M}_{(j)}$ and $\mathcal{%
M}_{(j)}^{0}$ can be decomposed with respect to $O(4)$ angular momenta.
Thus, we can separate the operators into partial waves, which can be written
as radial operators in the following form%
\begin{eqnarray}
\mathcal{M}_{(j,l)} &=&-\frac{d^{2}}{d\rho ^{2}}-\frac{N}{\rho }\frac{d}{%
d\rho }+\frac{l(l+2)}{\rho ^{2}}+\mathcal{V}_{j+1}(\rho ),  \label{sp1} \\
\mathcal{M}_{(j,l)}^{0} &=&-\frac{d^{2}}{d\rho ^{2}}-\frac{N}{\rho }\frac{d}{%
d\rho }+\frac{l(l+2)}{\rho ^{2}}+\mathcal{V}_{j}(\rho ),  \label{sp2}
\end{eqnarray}
}
\noindent \textcolor{black}{where $\mathcal{V}_{j}(\rho )\equiv V_{j}^{{\prime \prime }}(\phi
^{(j)})$. For simplicity, let us define%
\begin{equation}
G_{(j,l)}\equiv \frac{\det (\mathcal{M}_{(j,l)})}{\det (\mathcal{M}%
_{(j,l)}^{0})}.  \label{sp3}
\end{equation}
}
\textcolor{black}{In this case, $G_{(j,l)}$ can be computed by the so-called Gelfand-Yaglom
(G-Y) method \cite{gelfand}, which efficiently and allows to find the
determinant of an ordinary differential operator without necessity of
compute its eigenvalues. The G-Y method states that for the radial operators
(\ref{sp1}) and (\ref{sp2}), $G_{(j,l)}$ can be easily computed in the
following form%
\begin{equation}
G_{(j,l)}=\left[ \lim_{\rho \rightarrow \infty }\frac{\Psi _{(j,l)}(\rho )}{%
\Psi _{(j,l)}^{0}(\rho )}\right] ^{(l+1)^{2}},  \label{sp4}
\end{equation}
}
\noindent\textcolor{black}{where $\Psi _{(j,l)}(\rho )$ and $\Psi _{(j,l)}^{0}(\rho )$ are
regular solutions of the equations%
\begin{equation}
\mathcal{M}_{(j,l)}\Psi _{(j,l)}(\rho )=0,\text{ \ \ }\mathcal{M}%
_{(j,l)}^{0}\Psi _{(j,l)}^{0}(\rho )=0.  \label{sp5}
\end{equation}
}
\textcolor{black}{Here, it is important to highlight that there are three important types of
eigenvalue associated to the fluctuation operator. The first is the $l=0$ sector, which
has  a negative eigenvalue mode of the fluctuation operator. In this case,
it is responsible for the instability of the configuration and leads to
decay. The second is the $l=1$ sector, where there is a four-fold degenerate
zero eigenvalue (zero modes) of the fluctuation operator. From a physical
viewpoint, these four zero modes are equivalent to the Goldstone modes,
that exist due to breaking of translational invariance. Finally, the third type consists of the $l\geq 2$ sectors, with positive eigenvalues. However, we emphasize
that in our analysis the zero modes are removed from the fluctuation determinant. }

{Our next step, in order to compute the decay rate, is to compute the
explicit form of the Euclidean action. To accomplish it, it is necessary to
determine the integral (\ref{rel}) for each region. Let us then write $\rho
=\zeta /\sqrt{\lambda }_{j}$, in such a way that Eq. (\ref{9.11}) reads 
\begin{equation}
\phi _{j}(\zeta )=-{A}_{j}+\zeta ^{\nu }[\tilde{a}_{j}K_{\nu }(\zeta )+%
\tilde{b}_{j}I_{\nu }(\zeta )],  \label{rel1}
\end{equation}%
}\noindent 
\textcolor{black}{where $\tilde{a}_{j}\equiv a_{j}/\sqrt{\lambda _{j}}$, and $\tilde{b}_{j}\equiv b_{j}/\sqrt{\lambda _{j}}$.
In this case, Eq. (\ref{rel}) can  be rewritten as\begin{equation}
S_{E}^{(j)}=d_{(j,\nu )}\!\int_{\zeta _{j-1}/\sqrt{\lambda _{j}}}^{\zeta _{j}/\sqrt{\lambda _{j}}}\zeta ^{2\nu +1}\left[ \frac{1}{2}\left( \frac{d\phi _{j}}{d\zeta }\right) ^{2}\!\!+V_{j}[\phi _{j}(\zeta )]\right] d\zeta ,
\end{equation}
} \noindent 
\textcolor{black}{with the following redefinitions\begin{eqnarray}
&&\left. d_{(j,\nu )}:=\frac{2\pi ^{\nu +1}}{\lambda _{j}^{\nu }\Gamma (\nu
+1)},\right.  \label{rel2} \\
&&\left. V_{j}[\phi _{j}(\zeta )]:=\frac{\lambda _{j}}{2}\left[ \left( \phi
_{j}(\zeta )+G_{j}\right) ^{2}+b_{j}\right] ,\right.  \label{rel3}
\end{eqnarray}
} \noindent 
\textcolor{black}{where $G_{j}$ corresponds to the value in each vacuum. In addition,
the derivative $d\phi _{j}/d\zeta $ is crucial in our calculation, yielding 
\begin{equation}
\frac{d\phi _{j}}{d\zeta }=\zeta ^{\nu }[-\tilde{a}_{j}K_{\nu -1}(\zeta )+\tilde{b}_{j}I_{\nu -1}(\zeta )].
\end{equation}
} 
\textcolor{black}{Other important results to find the complete Euclidean action form are the
integrals \begin{eqnarray}
&&\left. \int d\zeta \zeta ^{m}K_{\nu }(\zeta )=2^{-(\nu +2)}\pi \zeta
^{m-\nu +1}\csc (\pi \nu )\right.   \notag \\
&&\left. \times \left\{ \left[ 4^{\nu }\Gamma \left( \frac{m-\nu +1}{2}\right) -\zeta ^{2\nu }\Gamma \left( \frac{m+\nu +1}{2}\right) \right]
\right. \right.   \notag \\
&&\left. \left. \times \sum_{l=1}^{2}\mathcal{H}_{RZ}\left[ \left\{
Q_{l}\right\} ,\left\{ \mathcal{R}_{l},\mathcal{T}_{l}\right\} ,\frac{\zeta
^{2}}{4}\right] \right\} ,\right. 
\end{eqnarray}
} \noindent 
\textcolor{black}{and\begin{eqnarray}
&&\left. \int d\zeta \zeta ^{m}I_{\nu }(\zeta )=2^{-(\nu +2)}\zeta ^{m-\nu
+1}\Gamma \left( \frac{m+\nu +1}{2}\right) \right.   \notag \\
&&\left. \times \mathcal{H}_{RZ}\left[ \left\{ Q_{2}\right\} ,\left\{ 
\mathcal{R}_{2},\mathcal{T}_{2}\right\} ,\frac{\zeta ^{2}}{4}\right]
,\right. 
\end{eqnarray}
} \noindent 
\textcolor{black}{where\begin{eqnarray}
Q_{l}\equiv\frac{m\!+\!(-\nu )^{l}\!+\!1}{2},\text{ }\,\,\mathcal{R}_{l}\equiv
1+(-\nu)^{l},\text{ }\,\,
\mathcal{T}_{l} =Q_{l}+1.  \notag
\end{eqnarray}
} 
\textcolor{black}{Furthermore, the function $\mathcal{H}_{RZ}\left[ \left\{ Q_{l}\right\}
,\left\{ \mathcal{R}_{l},\mathcal{T}_{l}\right\} ,\zeta ^{2}/4\right] $ is
the regularized generalized hypergeometric function and $\Gamma (X)$ is the
gamma function. Thus, by using of the above relations and} the usual
proceeding presented in \cite{dunne/2006, dunne/2008}, we can obtain, after %
\textcolor{black}{straightforward calculations}, the value of the
fluctuation determinant (\ref{dqg}) in each region, and as a consequence the
decay rates. In this case, the transition rates can be determined from a
straightforward example. For $\nu =1$, $\lambda _{1}=1$, $a=5$, $b_{1}=1$, $%
b_{2}=0.5$, $b_{3}=0.1$, $b_{4}=0$, $\rho _{1}=10^{-3}$, $\rho _{2}=10^{-2}$
and $\rho _{3}=10^{-1}$, we find that 
\begin{align}
\Gamma _{1}/V_{1}& \sim 10^{-25},\text{ } \\
\Gamma _{2}/V_{2}& \sim 10^{-14}, \\
\Gamma _{3}/V_{3}& \sim 10^{-5}.
\end{align}

At this point, it is important to remark that here the advantage is that in
each region the model is exactly solvable. Hence the decay rate was obtained
in an analytical form. 
\textcolor{black}{Another important consequence is
that the approach used can be applied to study the phase transitions in
models where we have piecewise potentials.}

\section{Concluding Remarks and Outlook}

\label{cro}

We investigated a class of oscillating configurations for the evolution of
the domain walls in Euclidean space. We found in an analytical form the
configurations of the field and the decay rate of the false vacuum.

Based on cosmological numerical values we can further incorporate a better
description of the phase transitions, as follows. In the cosmological
context, the Universe is known to have different dominant dynamical
components since its origin. From radiation and relativistic matter at early
time, to late time dark energy component, passing through the
matter-dominated scenario. In further work we intend to physically interpret
the vacuum decays of our model as the transition of different dynamical eras
of the universe, i.e., the transition from a radiation to a matter-dominated
scenario will be analysed merely as a transition from a false to a true
vacuum and the same will happen for the other dynamical transitions of the
universe. To accomplish it, we will need the values for $\vert\vec{x}\vert$
and $\tau$ for each transition era of the Universe.

From Fig.\ref{Figure2}, it is possible to perceive our argumentation. One
can interpret each of the regions $I$, $II$, $III$ and $IV$ as the different
dynamical eras of the Universe. While regions $II$, $III$ and $IV$ would
stand for the radiation, matter and dark energy-dominated eras, the region $%
I $ could account for the inflationary era (check, for instance, \cite%
{guth/1981,linde/1982,peebles/1999,chung/2007}). Note that the scalar field
evolution also makes one able to predict the universe fate, through a deep
analysis on region $IV$. Such an analysis may, for example, predicts the
universe to expand forever, yielding a Big-Freeze model \cite{yurov/2008}.

The transition rates obtained offer a further interpretation for the
dynamical evolution of the Universe. In fact, even if the state of the early
Universe was cold enough not to provide a thermal transition to the true
vacuum state, a quantum decay from the false vacuum to the true vacuum may
still be possible through the barrier mechanism.

\acknowledgments
RACC thanks to UFABC and CAPES for financial support. RdR thanks to CNPq
grants No. 303027/2012-6 and No. 473326/2013-2 for partial financial support, and to FAPESP grant No. 2015/10270-0.

\end{document}